\documentclass[12pt]{article}
\usepackage{a4wide}
\usepackage{amssymb}
\usepackage{hyperref}
\begin{document}
{\renewcommand{\thefootnote}{\fnsymbol{footnote}}
\begin{center}
{\LARGE  Minisuperspace models as infrared contributions}\\
\vspace{1.5em}
Martin Bojowald\footnote{e-mail address: {\tt bojowald@gravity.psu.edu}}
and Suddhasattwa Brahma\footnote{e-mail address: {\tt sxb1012@psu.edu}}
\\
\vspace{0.5em}
Institute for Gravitation and the Cosmos,\\
The Pennsylvania State
University,\\
104 Davey Lab, University Park, PA 16802, USA\\
\vspace{1.5em}
\end{center}
}

\setcounter{footnote}{0}

\begin{abstract}
  A direct correspondence of quantum mechanics as a minisuperspace model for a
  self-interacting scalar quantum-field theory is established by computing, in
  several models, the infrared contributions to 1-loop effective potentials of
  Coleman--Weinberg type. A minisuperspace {\em approximation} rather than
  truncation is thereby obtained. By this approximation, the spatial averaging
  scale of minisuperspace models is identified with an infrared scale (but not
  a regulator or cut-off) delimiting the modes included in the minisuperspace
  model.  Some versions of the models studied here have discrete space or
  modifications of the Hamiltonian expected from proposals of loop quantum
  gravity. They shed light on the question of how minisuperspace models of
  quantum cosmology can capture features of full quantum gravity. While it is
  shown that modifications of the Hamiltonian can well be described by
  minisuperspace truncations, some related phenomena such as signature change,
  confirmed and clarified here for modified scalar field theories, require at
  least a perturbative treatment of inhomogeneity beyond a strict
  minisuperspace model. The new methods suggest a systematic
  extension of minisuperspace models by a canonical effective formulation of
  perturbative inhomogeneity.
\end{abstract}

\section{Introduction}

Quantum mechanics can be seen as a ``minisuperspace model'' of quantum-field
theory obtained by restricting the latter to spatially constant
fields. Standard derivations in these frameworks, interpreted appropriately,
can therefore shed light on the question of how minisuperspace models of
quantum cosmology \cite{DeWitt} might be related to some putative full theory
of quantum gravity, and what physical information they can be able to
capture. Here, we initiate a detailed treatment of this form by comparing
semiclassical contributions to effective potentials in quantum mechanics with
different versions of the Coleman--Weinberg potential of self-interacting
scalar quantum-field theories \cite{ColemanWeinberg}.

At first sight, these two potentials look very different from each other,
suggesting that a relationship between the quantum-mechanical result, as our
minisuperspace model, and full quantum-field theory may not be obvious. A
quantum-mechanical system with Hamiltonian $\hat{H}_{\rm
  QM}=\frac{1}{2}\hat{p}^2+V(\hat{q})$ has, to first order in $\hbar$, a
semiclassical contribution to its effective potential given by
\begin{equation} \label{Veff}
 V_{\rm eff}(q)=V(q)+ \frac{1}{2}\hbar \sqrt{V_{qq}(q)}\,,
\end{equation}
as it can be derived by path-integral methods \cite{EffAcQM} or canonically
\cite{EffAc}. (Subscripts indicate the order of derivatives by the argument
of a function, here the potential.)

The Coleman--Weinberg potential, on the other hand,
can be expressed in various versions none of which suggest a clear
comparison. With a covariant cut-off and a quartic potential in the scalar
field $\phi(x)$, the renormalized potential given in \cite{ColemanWeinberg} is
\begin{equation}
 W_{\rm renorm}(\phi_0)= \lambda \phi_0^4 +\frac{9\lambda^2}{4\pi^2}
 \phi_0^4 \left(\log(\phi_0^2/M^2) -25/6\right)
\end{equation}
with the renormalization scale $M$ and $\phi_0$ the spatially constant
background value chosen for $\phi(x)$. Before a cut-off and renormalization
are employed, the potential is expressed as an integral over modes of the
quantum field,
\begin{equation} \label{CW}
 W_{\rm eff}(\phi_0) = \lambda \phi_0^4+ \frac{1}{2}i\hbar \int
 \frac{{\rm d}^4k}{(2\pi)^4} \log \left(1+
   \frac{12\lambda\phi_0^2}{||{\bf k}||^2}\right)
\end{equation}
or, performing the integration over the time component $k^0$,
\begin{equation} \label{CWCan}
  W_{\rm eff}(\phi_0) = \lambda \phi_0^4+ \frac{1}{2}\hbar \int
 \frac{{\rm d}^3k}{(2\pi)^3}  \left(\sqrt{|\vec{k}|^2+12\lambda
     \phi_0^2}- |\vec{k}|\right)\,.
\end{equation}

In this last form, the effective potential is directly obtained by canonical
methods \cite{CW}, briefly reviewed below. Being canonical, the derivation
does not make covariance manifest, and accordingly only the spatial wave
vector $\vec{k}$ appears. The formal derivation from (\ref{CW}) indeed shows
that one cannot directly use a covariant cutoff in this setting because $k^0$
must be integrated over its whole infinite range for (\ref{CWCan}) to
result. The remaining integral is still divergent, so that it can only be
treated by a non-covariant cut-off limiting the range of $|\vec{k}|^2$, rather
than the range of $||{\bf k}||^2$. Nevertheless, the expression (\ref{CWCan})
has advantages in our present context because it makes a comparison with
(\ref{Veff}) more clear. Indeed, just as (\ref{Veff}), the square root in
(\ref{CWCan}) contains the term $12\lambda\phi_0^2=W_{\phi\phi}(\phi_0)$, the
second derivative of the self-interacting potential $W(\phi)=\lambda\phi^4$ of
the Coleman--Weinberg model.

As we will demonstrate in what follows, the quantum-mechanical effective
potential can be extracted as the infrared contribution to the
Coleman--Weinberg potential of the corresponding quantum-field theory. In this
way, a direct relation between a minisuperspace model and its full theory can
be established, providing useful insights about the related question in
quantum gravity, where not much is known about concrete results in a possible
full theory. We will also be led to a systematic formulation of a
minisuperspace approximation as an expansion of the infrared contribution by
orders of $k_{\rm IR}/\sqrt{W_{\phi\phi}(\phi_0)}$ with the wave number
$k_{\rm IR}$ corresponding to the infrared scale. Perturbative inhomogeneity
will be shown to combine quantum mechanics as a minisuperspace model with a
quantum-field theory of modes coupled to it, providing a manageable setting
situated between a strict minisuperspace model and the corresponding full
theory.

\section{Quantum mechanics as a minisuperspace model}

We first have to make the minisuperspace picture of quantum mechanics more
precise. It should be a model of quantum-field theory seen as the full theory,
obtained by performing a spatial averaging of the fields. The traditional
construction of minisuperspace models starts at the classical level where the
averaging is performed, and then applies quantization methods to the resulting
model. The relationship between the quantum-mechanics model constructed in
this way and the full quantum-field theory has remained obscure. We will
arrive at a clear link by comparing effective potentials in both settings.

\subsection{Minisuperspace model}

Our classical full theory is a scalar field theory on Minkowski space-time
with Lagrangian
\begin{equation}
 L=\int{\rm d}^3x \left(\frac{1}{2}\dot{\phi}^2-\frac{1}{2}|\nabla\phi|^2
-W(\phi)\right)
\end{equation}
with some potential $W(\phi)$. The integral is performed over all of space up
to infinity. In order to obtain a finite Lagrangian after averaging the field
(setting it to a spatially constant value $\phi_0$), we have to choose a
bounded region of coordinate volume $\int{\rm d}^3x=V_0$, so that the
averaged, minisuperspace Lagrangian is
\begin{equation}
 L_{\rm mini}= V_0 \left(\frac{1}{2}\dot{\phi}^2-W(\phi)\right)\,.
\end{equation}
(At the minisuperspace level, we drop the subscript ``$0$'' of $\phi$ or its
momentum introduced below.)  The choice of the integration region and its
volume does not matter for the classical theory because we can evaluate a
constant field anywhere we like. However, as we will see now, the value of
$V_0$ does affect some quantum properties, which has been a constant source of
puzzlement in recent investigations of (especially loop) quantum cosmology.

We switch to the Hamiltonian picture by introducing the momentum $p=\partial
L_{\rm mini}/\partial\dot{\phi}=V_0\dot{\phi}$. The classical Hamiltonian is
\begin{equation}
 H_{\rm mini}= \frac{1}{2}\frac{p^2}{V_0}+V_0W(\phi)\,.
\end{equation}
At this stage, we can easily quantize the minisuperspace model, giving us the
Hamiltonian operator
\begin{equation}
  \hat{H}_{\rm mini} = \frac{1}{2}\frac{\hat{p}^2}{V_0}+V_0W(\hat{\phi})\,.
\end{equation}
Our minisuperspace model differs from standard quantum mechanics only in the
presence of coefficients depending on $V_0$. Nevertheless, the usual
methods can, of course, be applied.

We use the canonical effective methods of \cite{EffAc,Karpacz} and derive the
effective Hamiltonian by expanding the expectation value $\langle\hat{H}_{\rm
  mini}\rangle$ in a generic semiclassical state in terms of moments of the
state. We will only use second-order moments, appropriate for a semiclassical
expansion to first order in $\hbar$: We have the two fluctuations
$\Delta(\phi^2)=(\Delta\phi)^2$ and $\Delta(p^2)=(\Delta p)^2$, as well as the
covariance $\Delta(\phi
p)=\frac{1}{2}\langle\hat{\phi}\hat{p}+\hat{p}\hat{\phi}\rangle- \phi p$. Here
and in what follows, we simplify the notation by dropping brackets around
expectation values of basic operators, $\langle\hat{\phi}\rangle=\phi$ and
$\langle\hat{p}\rangle=p$.

Following this procedure, we obtain the effective Hamiltonian
\begin{equation} \label{Heff}
 H_{\rm eff} = \frac{1}{2}\frac{p^2}{V_0}+V_0W(\phi)
 +\frac{1}{2V_0} \Delta(p^2)+ \frac{1}{2} V_0 W_{\phi\phi}(\phi)
 \Delta(\phi^2)+\cdots
\end{equation}
where the dots indicate higher-moment terms, resulting from an expansion of
the potential. The effective Hamiltonian provides Hamiltonian equations of
motion for expectation values and moments, which can be derived from a Poisson
bracket defined using commutators:
$\{\langle\hat{A}\rangle,\langle\hat{B}\rangle\}:=
\langle[\hat{A},\hat{B}]\rangle/i\hbar$ (extended to products of expectation
values by the Leibnitz rule). For second-order moments, we obtain
\begin{eqnarray}
 \dot{\Delta}(\phi^2) &=& \frac{2}{V_0} \Delta(\phi p) \label{Deltaphidot}\\
 \dot{\Delta}(\phi p) &=& \frac{1}{V_0} \Delta(p^2)-
 V_0W_{\phi\phi}(\phi)\Delta(\phi^2)\\
 \dot{\Delta}(p^2) &=& -2V_0 W_{\phi\phi}(\phi)\Delta(\phi p)\,. \label{Deltapdot}
\end{eqnarray}

The effective potential is well-defined if the moments behave adiabatically,
corresponding to a derivative expansion in time \cite{HigherTime}. To lowest
order in an adiabatic expansion, we ignore the time derivatives on the
left-hand sides of (\ref{Deltaphidot})--(\ref{Deltapdot}) and solve the
resulting linear equations for the moments. We obtain $\Delta_0(\phi p)=0$ and
$\Delta_0(p^2)=V_0^2W_{\phi\phi}(\phi)\Delta_0(\phi^2)$. Saturating the
uncertainty relation $\Delta(\phi^2)\Delta(p^2)-\Delta(\phi p)^2\geq
\hbar^2/4$ in order to minimize fluctuations, we determine all the
second-order moments:
\begin{equation}
 \Delta_0(\phi^2)= \frac{1}{2}\frac{\hbar}{V_0\sqrt{W_{\phi\phi}(\phi)}}
 \quad,\quad
 \Delta_0(p^2) = \frac{1}{2} \hbar V_0\sqrt{W_{\phi\phi}(\phi)}\,.
\end{equation}
The effective potential in (\ref{Heff}) can then be expressed solely in terms
of $\phi$:
\begin{equation}
 V_{\rm eff}(\phi) = V_0W(\phi)
 +\frac{1}{2V_0} \Delta_0(p^2)+ \frac{1}{2} V_0 W_{\phi\phi}(\phi)
 \Delta_0(\phi^2)=
 V_0W(\phi)+ \frac{1}{2}\hbar \sqrt{W_{\phi\phi}(\phi)}\,.
\end{equation}

In order to arrive at this result of the effective potential expressed solely
in terms of $\phi$, eliminating the moments in (\ref{Heff}), two assumptions
were necessary: The validity of an adiabatic approximation for the moments and
the saturation condition of uncertainty relations. Both assumptions have been
shown to lead to correct results for anharmonic oscillators in quantum
mechanics \cite{EffAc,HigherTime} and, in the context of the Coleman--Weinberg
potential, for quartic self-interactions of a scalar field \cite{CW}. In the
rest of this paper, we work with general potentials $W(\phi)$ but assume that
approximations as in the preceding brief derivation are valid both in the
(full) quantum-field theory and the minisuperspace quantum-mechanics
model. Our main interest here lies in relating these two frameworks, not in
probing the range of semiclassical methods.

We divide by $V_0$ in order to extract the effective version of the original
potential $W(\phi)$ of the scalar theory before averaging:
\begin{equation} \label{Weff}
 W_{\rm eff}(\phi) = W(\phi)+ \frac{\hbar}{2V_0}\sqrt{W_{\phi\phi}(\phi)}\,.
\end{equation}
As promised, while the classical potential does not depend on the averaging
volume $V_0$, its first-order quantum correction does. The meaning of this
parameter cannot be understood within the minisuperspace model. The classical
reduction would suggest that the value of $V_0$ should not play any role in
physical results because the physics of an exactly constant field should not
depend on the volume of the region in which the field is constant, provided
the theory is local. Indeed, results in the classical model do not depend on
the value of $V_0$. Quantum effects, however, are known to be non-local, which
might explain the presence of $V_0$ in the correction term of
(\ref{Weff}). Still, this general observation does not elucidate the physical
meaning of $V_0$. In order to understand this issue, we now relate the
minisuperspace result to a calculation in the corresponding full theory.

\subsection{Effective potential in the infrared}

If we consider the integral in (\ref{CWCan}) only for small values of $k$,
that is in the infrared, we obtain an effective potential of the form
(\ref{Weff}), up to numerical factors. Indeed, if we include modes with
$|\vec{k}|$ between zero and $2\pi/R_0$, with a large $R_0$ for the maximum
wave length allowed, we can approximate the integral as the volume
$\frac{4}{3}\pi (2\pi/R_0)^3$ times the integrand evaluated at
$|\vec{k}|=0$. The result, $\frac{2}{3}\pi \hbar
\sqrt{W_{\phi\phi}(\phi_0)}/R_0^3$, agrees with (\ref{Weff}) up to a numerical
factor if we identify $\frac{4}{3}\pi R_0^3$ with the averaging volume $V_0$.

More accurately, we can write the infrared contribution to the effective
potential as
\begin{eqnarray} \label{CWIR}
  W_{\rm IR}(\phi_0) &=& W(\phi_0)+ \frac{1}{4\pi^2}\hbar
  \int_0^{2\pi/R_0} {\rm d}k k^2 \left(\sqrt{k^2+W_{\phi\phi}(\phi_0)}- k\right)\\
&=& W(\phi_0)+ \frac{1}{32\pi^2}\hbar
\left(\sqrt{(2\pi/R_0)^2+W_{\phi\phi}(\phi_0)}
  \left(2 \left(\frac{2\pi}{R_0}\right)^3+ \frac{2\pi
      W_{\phi\phi}(\phi_0)}{R_0}\right)\right.\\
 &&\qquad\quad\left.- 2\left(\frac{2\pi}{R_0}\right)^4-
  W_{\phi\phi}(\phi_0)^2 \log
  \frac{\sqrt{(2\pi/R_0)^2+W_{\phi\phi}(\phi_0)}+
2\pi/R_0}{\sqrt{W_{\phi\phi}(\phi_0)}}
\right)\nonumber\\
&\approx&  W(\phi_0)+ \frac{5}{2^9\pi^2}\hbar
\frac{\sqrt{W_{\phi\phi}(\phi_0)}}{R_0^3}\,.
\end{eqnarray}
As before, identifying $V_0=\frac{4}{3}\pi R_0^3$ gives us (\ref{Weff}) up to
numerical factors.  In the last step, we have kept only the leading term in an
expansion by $\left(R_0\sqrt{W_{\phi\phi}(\phi_0)}\right)^{-1}$, so that we
require the second derivative of the potential to be much larger than the
largest wave number squared, $W_{\phi\phi}(\phi_0)\gg R_0^{-2}$. This
approximation means that we restrict $\phi_0$ to values such that the
potential dominates the spatial-derivative term in the field-theory
Hamiltonian: In
\begin{equation}
  \frac{1}{2}\nabla^2\phi+W(\phi)\approx
  \frac{1}{2}\left(-k^2+W_{\phi\phi}(\phi_0) \right)
    \delta\phi^2\,,
\end{equation}
using an expansion $\phi(x)=\phi_0+\delta\phi(x)$ around the minimum $\phi_0$
of $W(\phi)$, the potential term is then dominant.

\section{Infrared contributions to 1-dimensional field theories}

The results presented so far show that the low-energy effective potential in
quantum mechanics agrees, up to numerical factors, with the infrared
contribution to a related quantum-field theory. The averaging volume $V_0$ of
the minisuperspace model corresponds to the infrared scale used in the
quantum-field theory. In this section, we explore how different choices of
full theories with the same potential, which therefore produce the same
minisuperspace result, affect the infrared contribution. Although we focus on
effective potentials, similar correspondences can be established between
2-point functions of quantum-field theories and quantum-mechanical
fluctuations.

\subsection{1-dimensional field theory}

We begin by providing a more-detailed derivation of the field-theory effective
potential with a 1-dimensional spatial manifold, following \cite{CW}. We take
this opportunity to show more details of the derivation that leads to the
integral used in (\ref{CWIR}), but at the same time give an example which
demonstrates the dependence of the infrared contribution on the spatial
dimension.

The quantum Hamiltonian, formally expanded to second-order moments, is
\begin{equation}
 H_{\rm Q}=\frac{1}{2} \int{\rm d}x \left(\pi(x)^2+\phi'(x)^2+2W(\phi(x))+
   G^{2,0}(x,x)+ {\rm D}^2 G^{0,2}(x,x)+ W_{\phi\phi}(\phi(x)) G^{0,2}(x,x)\right)\,.
\end{equation}
The derivative ${\rm D}^2G^{0,2}(x,x)$ is defined as $\lim_{y\to x}{\rm
  d}^2G^{0,2}(x,y)/{\rm d}x{\rm d}y$. For the moments, we have equations of
motion
\begin{eqnarray}
 \dot{G}^{0,2}(y,z) &=& G^{1,1}(y,z)+G^{1,1}(z,y) \\
 \dot{G}^{1,1}(y,z) &=& G^{2,0}(y,z)- \left(W_{\phi\phi}(\phi)- \frac{{\rm d}^2}{{\rm
       d}y^2}\right) G^{0,2}(y,z) \label{G11dot}\\
 \dot{G}^{2,0}(y,z) &=& -\left(W_{\phi\phi}(\phi)- \frac{{\rm d}^2}{{\rm
       d}z^2}\right) G^{1,1}(y,z)- \left(W_{\phi\phi}(\phi)- \frac{{\rm d}^2}{{\rm
       d}y^2}\right) G^{1,1}(z,y) \label{G20SC}\,.
\end{eqnarray}
To leading order in an adiabatic expansion, the first equation implies
$G^{1,1}(z,y)=-G^{1,1}(y,z)$, upon which the last one implies that $G^{1,1}=0$
using standard boundary conditions.

We solve (\ref{G11dot}) by using a Fourier decomposition
\begin{equation}
 G^{0,2}(x,y) = \int {\rm d}k_x{\rm d}k_y f(k_x,k_y) e^{i(k_xx+k_yy)}\,.
\end{equation}
For a translation-invariant theory, $f(k_y,k_y)$ must be of the form
$g(k)\delta(k_x+k_y)$, so that
\begin{equation} \label{G02F}
 G^{0,2}(x,y) = \int {\rm d}k g(k) e^{ik(x-y)}\,.
\end{equation}
By (\ref{G11dot}), we then have
\begin{equation} \label{G20F}
 G^{2,0}(x,y) = \int {\rm d}k (W_{\phi\phi}(\phi)+k^2) g(k) e^{ik(x-y)}\,.
\end{equation}

The moments $G^{0,2}$ and $G^{2,0}$ (with $G^{1,1}=0$) appear in the
uncertainty relation
\begin{equation}
 G^{0,2}(x_1,x_2)G^{2,0}(y_1,y_2) \geq \frac{\hbar^2}{8}
 \left(\delta(x_1-y_1)\delta(x_2-y_2)+ \delta(x_1-y_2)\delta(x_2-y_1)\right)\,.
\end{equation}
The saturation condition is singular for moments such as (\ref{G02F}) and
(\ref{G20F}), but it can nevertheless be evaluated in order to restrict the
values. If we set $x_1=x_2$ in the saturated uncertainty relation and
integrate over this value, we can see that the free function $g(k)$ must be
\begin{equation}
 g(k)=\frac{\hbar}{2\pi} \frac{1}{2\sqrt{W_{\phi\phi}(\phi)+k^2}}
\end{equation}
if the theory is reflection symmetric, $g(-k)=g(k)$.
Therefore,
\begin{eqnarray}
 G^{0,2}(x,y) &=& \frac{\hbar}{2\pi} \int {\rm d}k
 \frac{1}{2\sqrt{W_{\phi\phi}(\phi)+k^2}} e^{ik(x-y)}\\
G^{2,0}(x,y) &=& \frac{\hbar}{2\pi} \int {\rm d}k
 \frac{1}{2}\sqrt{W_{\phi\phi}(\phi)+k^2} e^{ik(x-y)}\,.
\end{eqnarray}
We insert these moments in the effective Hamiltonian and arrive at the
Coleman--Weinberg potential
\begin{eqnarray}
 W_{\rm eff}(\phi_0) &=& W(\phi_0)+ \frac{\hbar}{4\pi} \int{\rm d}k
 \left(\sqrt{W_{\phi\phi}(\phi_0)+k^2}-|k|\right)\nonumber\\
&=&W(\phi_0)+\frac{\hbar}{2\pi} \int_0^k
   {\rm d}k\left(\sqrt{W_{\phi\phi}(\phi_0)+k^2}-k\right)
\end{eqnarray}
with a spatially constant $\phi_0$.  (In the integrand, we subtract $|k|$ in
order to ensure that there is no contribution in the free case when
$W=0$. This term amounts to an infinite subtraction of the diverging moments
$G^{a,b}(x,x)$.)

The integral in $W_{\rm eff}$ can be evaluated explicitly, but we are
interested only in the infrared contribution given by modes up to some scale
$2\pi/L_0$. Using the additional assumption of potential dominance at the
chosen value of $\phi_0$, $W_{\phi\phi}(\phi_0)\gg (2\pi/L_0)^2$, we obtain
\begin{equation}
 W_{\rm IR}(\phi_0) = W(\phi_0)+ \hbar \frac{\sqrt{W_{\phi\phi}(\phi_0)}}{L_0}\,.
\end{equation}
The result differs from the minisuperspace model by a factor of 2.

\subsection{Discrete space}
\label{s:Disc}

We now modify our 1-dimensional field theory by putting it on a discrete
space. The classical Hamiltonian is
\begin{equation}
 H=\sum_{n=-\infty}^{\infty} \left(\frac{1}{2}\pi_n^2+ \frac{1}{8\ell_0^2}
   (\phi_{n+1}-\phi_{n-1})^2+ W(\phi_n)\right)\,.
\end{equation}
We have chosen a symmetric discretization of the spatial derivative, and
denoted the discreteness scale by $\ell_0$. Accordingly, there will be a
maximal wave number $k_{\rm max}=2\pi/\ell_0$ in Fourier decompositions.

We obtain the quantum Hamiltonian, as before to second-order moments, given by
\begin{eqnarray}
 H_{\rm Q} &=& \frac{1}{2}\sum_{n=-\infty}^{\infty} \left(\pi_n^2+ \frac{1}{4\ell_0^2}
   (\phi_{n+1}-\phi_{n-1})^2+ 2W(\phi_n)\right.\nonumber\\
 &&\left. +G_{n,n}^{2,0}+ \frac{1}{4\ell_0^2}
   (G_{n+1,n+1}^{0,2}- 2G_{n+1,n-1}^{0,2}+ G_{n-1,n-1}^{0,2})+W_{\phi\phi}(\phi_n)
   G_{n,n}^{0,2} \right)\,.
\end{eqnarray}
It is slightly easier to compute equations of motion for the moments if we
first rearrange the spatial difference term by a suitable shift of the
summation labels:
\begin{eqnarray}
 H_{\rm Q} &=& \frac{1}{2}\sum_{n=-\infty}^{\infty} \left(\pi_n^2+ \frac{1}{4\ell_0^2}
   (\phi_{n+1}-\phi_{n-1})^2+ 2W(\phi_n)\right.\nonumber\\
 &&\left. +G_{n,n}^{2,0}+ \frac{1}{2\ell_0^2}
   (G_{n,n}^{0,2}- G_{n+1,n-1}^{0,2})+W_{\phi\phi}(\phi_n)
   G_{n,n}^{0,2} \right)\,.
\end{eqnarray}

Equations of motion are
\begin{eqnarray}
  \dot{G}_{n,m}^{0,2} &=& G_{n,m}^{1,1}+G_{m,n}^{1,1}\\
  \dot{G}_{n,m}^{1,1} &=& G_{n,m}^{2,0}- W_{\phi\phi}(\phi_n) G_{n,m}^{0,2}-
  \frac{1}{4\ell_0^2}
  (2G_{n,m}^{0,2}-G_{n,m-2}^{0,2}-G_{n,m+2}^{0,2}) \label{G11dotdisc}\\
  \dot{G}_{n,m}^{2,0} &=& -\frac{1}{4\ell_0^2} (2G_{n,m}^{1,1}-G_{n-2,m}^{1,1}-
  G_{n+2,m}^{1,1}+ 2G_{m,n}^{1,1}- G_{m-2,n}^{1,1}- G_{m+2,n}^{1,1})\,.
\end{eqnarray}
In an adiabatic approximation, the first equation implies
$G_{n,m}^{1,1}=-G_{m,n}^{1,1}$, upon which the last equation gives
\begin{equation}
 G_{n-2,m}^{1,1}+G_{n+2,m}^{1,1} = G_{n,m-2}^{1,1}+G_{n,m+2}^{1,1}\,.
\end{equation}
Boundary conditions are now more complicated, but we may use
$G_{n,m}^{1,1}=0$ for the correct continuum limit of the preceding subsection
to be realized.

In order to solve (\ref{G11dotdisc}), we write
\begin{equation}
 G_{n,m}^{0,2} = \int_0^{k_{\rm max}}{\rm d}k g(k) e^{ik\ell_0(n-m)}
\end{equation}
as a Fourier decomposition in the require $k$-range. Equation
(\ref{G11dotdisc}) then implies
\begin{eqnarray}
 G_{n,m}^{2,0} &=& \left(W_{\phi\phi}(\phi_n)+\frac{1}{2\ell_0^2}\right) G_{n,m}^{0,2}-
 \frac{1}{4\ell_0^2} (G_{n,m-2}^{0,2}+G_{n,m+2}^{0,2})\\
 &=& \int_0^{k_{\rm max}} {\rm d}k g(k) \left(W_{\phi\phi}(\phi_n)
  - \frac{1}{4\ell_0^2}\left(e^{2ik\ell_0}+ e^{-2ik\ell_0}-2\right)\right)
e^{ik\ell_0(n-m)}\\
 &=& \int_0^{k_{\rm max}} {\rm d}k g(k) \left(W_{\phi\phi}(\phi_n)
  + \frac{\sin^2(\ell_0k)}{\ell_0^2}\right)e^{ik\ell_0(n-m)}\,.
\end{eqnarray}
We saturate the uncertainty relation in the form
$\sum_nG_{n,m}^{0,2}G_{n,m'}^{2,0}= \frac{1}{4}\hbar^2 \delta_{m,m'}$ and
obtain the effective potential
\begin{equation}
 W_{\rm eff}(\phi_0) = W(\phi_0)+ \frac{\hbar}{2\pi} \int_0^{k_{\rm max}}{\rm d}k
 \left(\sqrt{W_{\phi\phi}(\phi_0)+\sin^2(\ell_0k)/\ell_0^2}-
\sin(\ell_0k)/\ell_0\right)\,.
\end{equation}

The integral can again be evaluated. For an infrared scale of $2\pi/L_0\ll
2\pi/\ell_0$ and potential domination, $W_{\phi\phi}(\phi_0)\gg (2\pi/L_0)^2$,
we have
\begin{eqnarray}
 W_{\rm IR}(\phi_0) &=& W(\phi_0)
 + \hbar \left(\frac{\sqrt{W_{\phi\phi}(\phi_0)}}{L_0}+
   \frac{8}{3} \pi^2 \frac{1}{\sqrt{W_{\phi\phi}(\phi_0)}L_0^3}\right.\\
 &&\left.- \frac{2}{15} \pi^4
   \frac{4W_{\phi\phi}(\phi_0)\ell_0^2+3}{W_{\phi\phi}(\phi_0)^{3/2} L_0^5}+
   W_{\phi\phi}(\phi_0)
   O\left(L_0^{-6}W_{\phi\phi}(\phi_0)^{-3}\right)\right)\,.
 \nonumber
\end{eqnarray}
As the expansion to higher orders in
$\left(L_0\sqrt{W_{\phi\phi}(\phi_0)}\right)^{-1}$ shows, one has to go well
beyond the leading result of the infrared expansion in order to see the
discreteness scale $\ell_0$. Moreover, the fifth-order term depends
significantly on $\ell_0$ only if $W_{\phi\phi}(\phi_0)\ell_0^2 \approx 1$ or
larger, so that the second derivative of the potential must be huge for a
small discreteness scale.

\subsection{Periodic space}

In our next model, we assume that the spatial manifold is a circle with
coordinate radius ${\cal L}$. The Hamiltonian is
\begin{equation}
 H=\int_0^{\cal L}{\rm d}x \left(\frac{1}{2}
   \pi(x)^2+\frac{1}{2}\phi'(x)^2+W(\phi)\right)\,,
\end{equation}
and we impose periodic boundary conditions $\phi({\cal L})=\phi(0)$,
$\phi'({\cal L})=\phi'(0)$. All our local equations of motion are as in the
continuum model we started with, but Fourier decompositions are discrete. We
therefore write
\begin{equation}
 G^{0,2}(y,z)= \sum_{j=-\infty}^{\infty} g_j e^{2\pi i j(y-z)/{\cal L}}
\end{equation}
and the leading-order adiabatic equations imply that
\begin{equation}
 G^{2,0}(y,z)= \sum_{j=-\infty}^{\infty} g_j \left(W_{\phi\phi}(\phi)+
   \left(\frac{2\pi}{\cal L}\right)^2 j^2\right) e^{2\pi i j(y-z)/{\cal L}}\,.
\end{equation}
Upon saturating the uncertainty relation, the Fourier coefficients are
determined by
\begin{equation}
 g_j = \frac{\hbar}{\cal L} \frac{1}{2\sqrt{W_{\phi\phi}+(2\pi/{\cal L})^2j^2}}\,.
\end{equation}
The effective potential is
\begin{eqnarray} \label{Weffcompact}
 W_{\rm eff}(\phi_0) &=& W(\phi_0)+  \frac{\hbar}{2{\cal L}}
 \sum_{j=-\infty}^{\infty}
 \left(\sqrt{W_{\phi\phi}(\phi_0)+\left(\frac{2\pi}{\cal L}\right)^2 j^2}- \frac{2\pi
     j}{{\cal L}}\right)\nonumber\\
 &=& W(\phi_0)+ \frac{1}{2}\hbar \frac{\sqrt{W_{\phi\phi}(\phi_0)}}{\cal L}+
 \frac{\hbar}{\cal L} \sum_{j=1}^{\infty}
 \left(\sqrt{W_{\phi\phi}(\phi_0)+\left(\frac{2\pi}{\cal L}\right)^2 j^2}- \frac{2\pi
     j}{{\cal L}}\right)\,.
\end{eqnarray}

We choose a length $L_0\leq {\cal L}$ as the infrared scale which determines
the infrared contribution
\begin{equation}\label{WIRcompact}
 W_{\rm IR} = W(\phi_0)+ \frac{1}{2}\hbar \frac{\sqrt{W_{\phi\phi}(\phi_0)}}{\cal L}+
 \frac{\hbar}{\cal L} \sum_{j=1}^{{\cal L}/L_0-1}
 \left(\sqrt{W_{\phi\phi}(\phi_0)+\left(\frac{2\pi}{\cal L}\right)^2 j^2}- \frac{2\pi
     j}{{\cal L}}\right)\,.
\end{equation}
We have assumed that ${\cal L}/L_0=N$ is an integer, so that averaging over
the infrared scale eliminates all modes with $j=N$ or higher. The value $N=1$
is allowed, in which case the averaging is complete and only homogeneous modes
are left. There remains an infrared contribution to the classical potential,
given by
\begin{equation}
 W_{\rm IR}^{L_0={\cal L}}(\phi_0)=W(\phi_0)+\frac{1}{2}\hbar
 \frac{\sqrt{W_{\phi\phi}(\phi_0)}}{L_0}\,.
\end{equation}
In this case, the infrared contribution agrees exactly with the
quantum-mechanical effective potential (\ref{Weff}), even in its numerical
factor.

We are not restricted to $N=1$ as long as $N$ is a positive integer. For
$N>1$, the infrared contribution (\ref{WIRcompact}) is a finite sum with
contributions in addition to the quantum-mechanical one. If we assume
potential domination, as before, we have $W_{\phi\phi}(\phi_0)\gg (2\pi/L_0)^2\geq
(2\pi/{\cal L})^2$, and we can expand the square root in order to evaluate the
sums:
\begin{eqnarray}
 W_{\rm IR}(\phi_0) &=& W(\phi_0)+  \frac{1}{2}\hbar
 \frac{\sqrt{W_{\phi\phi}(\phi_0)}}{\cal
   L}\nonumber\\
&&+ \sum_{j=1}^{{\cal L}/L_0-1} \left(W_{\phi\phi}(\phi_0) \left(1+\frac{1}{2}
     \left(\frac{2\pi}{\cal L}\right)^2 \frac{1}{W_{\phi\phi}(\phi_0)}
     j^2+\cdots\right)-
   \frac{2\pi j}{\cal L}\right)\\
 &=& W(\phi_0)+  \frac{1}{2}\hbar \frac{\sqrt{W_{\phi\phi}(\phi_0)}}{\cal
   L}- \pi \frac{\hbar}{{\cal L}^2} \frac{\cal L}{L_0}\left(\frac{\cal
     L}{L_0}-1\right)\nonumber \\
&& +\frac{\pi^2}{3}  \frac{\hbar}{\sqrt{W_{\phi\phi}(\phi_0)}{\cal L}^3}
\frac{\cal
   L}{L_0}\left(\frac{\cal L}{L_0}-1\right)\left(2\frac{\cal
     L}{L_0}-1\right)+\cdots\,.
\end{eqnarray}
For large $N={\cal L}/L_0\gg 1$,
\begin{equation} \label{WIRCompact}
 W_{\rm IR}(\phi_0) = W(\phi_0)+  \frac{1}{2N}\hbar
 \frac{\sqrt{W_{\phi\phi}(\phi_0)}}{L_0}-
 \pi \frac{\hbar}{L_0^2}
 +\frac{\pi^2}{3}  \frac{\hbar}{\sqrt{W_{\phi\phi}(\phi_0)}L_0^3}\,.
\end{equation}

\subsection{Scalar field on a homogeneous metric background}

If the background space-time of the scalar field theory is not Minkowskian,
the spatial metric $q$ and lapse function $N$ (distinct from the mode number
in the preceding subsection) appear in the Hamiltonian
\begin{equation} \label{Hback}
 H=\int {\rm d}x N(x) \sqrt{q}\left( \frac{1}{2} \frac{\pi(x)^2}{q}+
   \frac{\phi'(x)^2}{q}+W(\phi)\right)\,.
\end{equation}
The momentum is now related to the time derivative of $\phi$ by
$\pi=\sqrt{q}\dot{\phi}$.  In one dimension the metric is just a function,
given by $q(x)=a^2$ for a homogeneous background with the cosmological scale
factor $a$.  For our purposes, the lapse function can be ignored, and we may
absorb the factor of $\sqrt{q}=a$ in the spatial coordinate $x$. The only
effect of the spatial metric is then to write the entire Hamiltonian,
including the spatial derivative of $\phi$, in terms of $ax$ instead of
$x$. Accordingly, mode expansions are formulated with respect to the physical
wave number $k/a$ instead of the comoving one, $k$. Similarly, if we use the
discrete model, the discreteness parameter in the spatial difference operator
is the geometrical distance $a\ell_0$ rather than the coordinate distance
$\ell_0$.

In the corresponding minisuperspace model the coordinate value $L_0$ of the
infrared scale is replaced by the geometrical distance $aL_0$ in all
equations, including the resulting effective potential. The dependence of the
effective potential on the infrared scale is therefore coordinate
invariant. It only depends on the physical scale used to average fields.

\subsection{Quantum-geometry modifications}

Models of loop quantum gravity have been used to suggest two types of
modifications which turn out to deform the classical notion of
covariance. Accordingly, the kinetic term of a scalar Hamiltonian is modified
by inserting different types of functions. These functions, in canonical form,
may depend on the spatial metric or extrinsic curvature, which in our context
would be treated as background fields, but they can also depend on the scalar
field or its momentum and then change the quadratic form of kinetic terms.
There are two important forms of modifications, inverse-triad corrections
first introduced in loop quantum cosmology in \cite{Inflation}, and holonomy
corrections introduced in \cite{EffHam,GenericBounce,AmbigConstr}.

\subsubsection{Inverse-triad corrections}

Inverse-triad corrections result from a quantization of the inverse metric
factors in a Hamiltonian such as (\ref{Hback}), in which the kinetic and
spatial-derivative term are divided by metric components. Loop quantizations
(using a densitized triad instead of the spatial metric) lead to discrete
metric operators with zero in their discrete spectra, so that no direct
inverse exists. Nevertheless, as proposed in \cite{QSDI,QSDV}, an inverse can
be quantized in an indirect way, schematically writing
$q^{-1/2}=2\{q^{1/2},p_q\}$ and replacing the Poisson bracket by a commutator
divided by $i\hbar$. A densely-defined operator results, but its spectrum
differs from the expected $q_n^{-1/2}$ for small values of $q_n$
\cite{InvScale,ICGC}.

One can model this effect of discrete quantum geometry by changing the
$q$-dependence of a Hamiltonian (\ref{Hback}) on a classical background.  With
two given positive functions $\nu$ and $\sigma$, a scalar Hamiltonian modified
by inverse-triad corrections has the form
\begin{equation} \label{Hinv}
 H=\int {\rm d}x \left(\frac{1}{2}\nu \pi(x)^2 + \frac{1}{2}\sigma \phi'(x)^2+
   W(\phi)\right)\,.
\end{equation}
For simplicity, we will assume $\nu$ and $\sigma$ to depend only on time,
perhaps implicitly, but not on $x$. The time dependence of the modification
functions could result from an expanding cosmological background, but we omit
corresponding factors of $q=a^2$.

We can eliminate the function $\sigma$ by employing a canonical transformation
$\tilde{\phi}:=\sqrt{\sigma}\phi$ and $\tilde{\pi}:=\pi/\sqrt{\sigma}$. The
only modification in the kinetic term is then a factor of $\beta:=\nu\sigma$
multiplying the momentum squared, and there is a substitution of
$\tilde{\phi}/\sqrt{\sigma}$ for $\phi$ in the potential. We will therefore
assume the Hamiltonian to be of the form
\begin{equation}
  H=\int {\rm d}x \left(\frac{1}{2}\beta \tilde{\pi}(x)^2 + \frac{1}{2}
    \tilde{\phi}'(x)^2+
    W(\tilde{\phi}/\sqrt{\sigma})\right)
\end{equation}
without changing our notation for the fields.

With our new Hamiltonian, the equation of motion for $G^{2,0}$, after
transitioning to an expanded quantum Hamiltonian, is unchanged, while
\begin{equation}
 \dot{G}^{0,2}(y,z) = \beta(G^{1,1}(y,z)+G^{1,1}(z,y))\,.
\end{equation}
To leading adiabatic order, the conclusion that $G^{1,1}=0$ is therefore
unchanged. For $G^{1,1}$, we have the equation of motion
\begin{equation}
  \dot{G}^{1,1}(y,z) = \beta G^{2,0}(y,z)-
  \left(\frac{W_{\phi\phi}(\tilde{\phi}/\sqrt{\sigma})}{\sigma}-
    \frac{{\rm d}^2}{{\rm d}y^2}\right) G^{0,2}(y,z)\,.
\end{equation}
Upon Fourier decomposition and saturating the uncertainty relation, we obtain
coefficients
\begin{equation}
 g(k) = \frac{\hbar}{2\pi}
 \frac{\sqrt{\beta}}{2\sqrt{W_{\phi\phi}(\tilde{\phi}/\sqrt{\sigma})/\sigma+ k^2}}\,.
\end{equation}

For potential domination, the infrared contribution to the effective potential
now reads
\begin{equation}
 W_{\rm IR}(\phi_0) = W(\tilde{\phi}_0/\sqrt{\sigma})+ \hbar
 \sqrt{\frac{\beta}{\sigma}}
 \frac{\sqrt{W_{\phi\phi}(\tilde{\phi}_0/\sqrt{\sigma})}}{L_0}=
W(\phi_0)+ \hbar
 \sqrt{\frac{\beta}{\sigma}}
 \frac{\sqrt{W_{\phi\phi}(\phi_0)}}{L_0}
\end{equation}
which differs from the quantum-mechanical result by an additional factor
$\sqrt{\beta/\sigma}=\sqrt{\nu}$ depending on the modification functions. The
same factor is obtained if one implements the modification in the
quantum-mechanical Hamiltonian
\begin{equation}
 \hat{H} = \frac{1}{2} \nu \hat{p}^2+W(\hat{q})
\end{equation}
and follows the derivations leading up to (\ref{Veff}).  The modification by
$\nu$ follows from a direct reduction of (\ref{Hinv}) to homogeneity.

\subsubsection{Holonomy corrections}

Holonomy modifications are motivated by the use of holonomies of a
gravitational connection as basic operators in loop quantum gravity, while
connection components cannot be quantized directly. Any object, such as a
Hamiltonian, which depends on the connection in its classical expression, must
therefore be modified for it to be quantized on the kinematical Hilbert space
of loop quantum gravity. The form of holonomies (or their matrix elements) of
compact groups suggests that polynomial dependences on the connection are
replaced by bounded and periodic functions.

\paragraph{Modified spatial-derivative term.}

Similarly, a scalar field is not represented directly on the kinematical
Hilbert space of loop quantum gravity, but only through its exponentials
$h_x(\phi):=\exp(i\phi(x))$. In our context, this modification would be
similar to inverse-triad corrections with $\nu=1$ and $\sigma$ related to a
derivative of the modification function.  However, in this case we can no
longer assume that the modification function depend only on time, so that a
more detailed analysis is required. (The final result will nevertheless turn
out to be closely related to the one for inverse-triad corrections because the
effective potentials to orders considered here depend on the field expectation
value only via its spatially constant average.)

In the presence of scalar holonomy modifications, the discrete model used in
Sec.~\ref{s:Disc} is more suitable. We now use a classical Hamiltonian
\begin{equation} \label{HScalarHol}
 H=\sum_{n=-\infty}^{\infty} \left(\frac{1}{2}\pi_n^2+ \frac{1}{4\ell_0^2}
   \left(g(\phi_n)^2-g(\phi_{n+1})g(\phi_{n-1})\right)+ W(g(\phi_n))\right)
\end{equation}
with some local function $g(\phi_n)$ which is non-linear in the presence of
scalar holonomy modifications. We compute the quantum Hamiltonian
\begin{eqnarray}
  H_{\rm Q} &=& \frac{1}{2} \sum_{n=-\infty}^{\infty} \left(\pi_n^2+
    \frac{1}{2\ell_0^2}    \left(g(\phi_n)^2-g(\phi_{n+1})g(\phi_{n-1})\right)+
    2W(g(\phi_n))\right.\\
  && \left.\qquad \qquad +G_{n,n}^{2,0}+ \frac{1}{2\ell_0^2} \left(
      \Delta_1g[\phi_n]
      G_{n,n}^{0,2}- \Delta_2g[\phi_n] G_{n+1,n-1}^{0,2}\right)+
    W_{\phi_n\phi_n}(g(\phi_n))
    G_{n,n}^{0,2}\right) \,, \nonumber
\end{eqnarray}
where we introduced the non-local functions
\begin{eqnarray}
 \Delta_1g[\phi_n] &:=& g_{\phi_n}(\phi_n)^2-\frac{1}{2} g_{\phi_n\phi_n}(\phi_n)
 \left(g(\phi_{n+2})- 2g(\phi_n)+ g(\phi_{n-1})\right)\\
 \Delta_2g[\phi_n] &:=& g_{\phi_{n+1}}(\phi_{n+1})g_{\phi_{n-1}}(\phi_{n-1})\,.
\end{eqnarray}
Both functions are equal to one in the unmodified case.

The equation of motion for $G_{n,m}^{0,2}$ is unchanged compared to the case
with $g(\phi_n)=\phi_n$, and while the equation for $G_{n,m}^{2,0}$ has
coefficients depending on $\Delta_1g[\phi_n]$ and $\Delta_2g[\phi_n]$, the
leading adiabatic solutions are still consistent with $G_{n,m}^{1,1}=0$. The
remaining equation is
\begin{equation}
 \dot{G}_{n,m}^{1,1} = G_{n,m}^{2,0}- \left(W_{\phi_n\phi_n}(g(\phi_n))+
   \frac{1}{2\ell_0^2} \Delta_1g[\phi_n]\right) G_{n,m}^{0,2}
 -\frac{1}{4\ell_0^2}\Delta_2g[\phi_n] (G_{n,m-2}^{0,2}+ G_{n,m+2}^{0,2})\,.
\end{equation}

Using the same Fourier decomposition as before, we solve this equation by
\begin{equation}
 G_{n,m}^{2,0} = \int_0^{k_{\rm max}} {\rm d}k g(k) \left( W_{\phi_n\phi_n}(g(\phi_n))+
   \frac{1}{2\ell_0^2} (\Delta_1g[\phi_n]-\Delta_2g[\phi_n])-
   \Delta_2g[\phi_n] \frac{\sin^2(\ell_0 k)}{\ell_0^2}\right)
 e^{ik\ell_0(n-m)}
\end{equation}
where $g(k)$ appears in
\begin{equation}
 G_{n,m}^{0,2} = \int_0^{k_{\rm max}} {\rm d}k g(k) e^{ik\ell_0(n-m)}\,.
\end{equation}
Saturating the uncertainty relation leads to the effective potential
\begin{equation}
 W_{\rm eff}(\phi_0) = W(g(\phi_0))+ \frac{\hbar}{2\pi} \int_0^{k_{\rm max}}{\rm d}k
 \left(\sqrt{W_{\phi\phi}(g(\phi_0))+ g_{\phi}(\phi_0)^2
     \sin^2(\ell_0k)/\ell_0^2}-\sin(\ell_0k)/\ell_0\right)\,.
\end{equation}
Here, we have used the fact that for constant $\phi_n=\phi_0$ (slightly
abusing the subscript notation), as assumed in effective potentials of
Coleman--Weinberg type,
$\Delta_1g[\phi_n]=g_{\phi}(\phi_0)^2=\Delta_2g[\phi_n]$. The subtraction is
chosen in such a way that it removes the contribution only when $W(\phi)=0$
and $g_{\phi}(\phi)=1$, because non-linear $g(\phi)$ imply non-quadratic
derivative terms in the Hamiltonian and therefore interactions even if the
classical potential vanishes.  The leading-order infrared contribution depends
on $g(\phi)$ only through the potential. The same modification is obtain in
the minisuperspace model of (\ref{HScalarHol}).

\paragraph{Modified momentum dependence.}

Alternatively, in order to model the formal dependence of gravitational
Hamiltonians on connection components, it is instructive to modify the
quadratic dependence of our scalar Hamiltonian on $\pi(x)$. The momentum of
the scalar field would then be represented by its exponentials of integrated
fields, $h_{x_0,\delta}(\pi):= \exp\left(\int_{x_0}^{x_0+\delta} {\rm d}x
  \pi(x)\right)$. (Written on a metric background, the momentum has a density
weight and can therefore be integrated directly without any additional measure
factors.) Using $h_{x_0,\delta}(\pi)$ instead of $\pi$ in the Hamiltonian
leads to a non-local expression, but as often done, here we model the effect
by replacing $\pi(x)^2$ by a local function $f(\pi(x))$:
\begin{equation}
 H=\int{\rm d}x \left(\frac{1}{2}f(\pi(x))+
   \frac{1}{2}\phi'(x)^2+W(\phi)\right)\,.
\end{equation}
In this form, the scalar model resembles the 3-dimensional version proposed in
\cite{PolyInfl,ExpScalar} and further evaluated in
\cite{ModUncert,PolyProp,PolyPrim,PolyInfl2}.

The quantum Hamiltonian
\begin{eqnarray}
 H_{\rm Q} &=& \frac{1}{2}\int{\rm d}x \Bigl(f(\pi(x))+\phi'(x)^2+2W(\phi)
 \nonumber\\
 &&\qquad+ f_{\pi\pi}(\pi)
   G^{2,0}(x,x)+ {\rm D}^2 G^{0,2}(x,x)+
   W_{\phi\phi}(\phi)G^{0,2}(x,x)\Bigr)\,,
\end{eqnarray}
expanded to second order in moments, implies equations of motion
\begin{equation}\label{G02SC}
 \dot{G}^{0,2}(y,z)=\frac{1}{2}f_{\pi\pi}(\pi) (G^{1,1}(y,z)+G^{1,1}(z,y))
\end{equation}
and an unchanged equation for $\dot{G}^{2,0}$, so that we conclude $G^{1,1}=0$
as before, using the leading adiabatic approximation. The remaining equation
to be solved is
\begin{equation}\label{G11SC}
 \dot{G}^{1,1}(y,z) = \frac{1}{2}f_{\pi\pi}(\pi) G^{2,0}(y,z)-
 \left(W_{\phi\phi}(\phi)-\frac{{\rm d}^2}{{\rm d}y^2}\right) G^{0,2}(y,z)\,.
\end{equation}
A Fourier decomposition can be done as before, just with additional factors of
$f_{\pi\pi}(\pi)/2$ in some solutions. We arrive at the effective
potential
\begin{equation} \label{WeffHol}
 W_{\rm eff}(\phi_0)=W(\phi_0)+ \frac{\hbar}{4\pi}
 \sqrt{\frac{f_{\pi\pi}(\pi_0)}{2}}
 \int{\rm d}k \left(\sqrt{W_{\phi\phi}(\phi_0)+k^2}-|k|\right)
\end{equation}
with a spatially constant $\pi_0$, and the infrared contribution is
\begin{equation} \label{WIRHol}
 W_{\rm IR}(\phi_0) = W(\phi_0)+ \hbar \sqrt{\frac{f_{\pi\pi}(\pi_0)}{2}}
 \frac{\sqrt{W_{\phi\phi}(\phi_0)}}{L_0}\,.
\end{equation}
The same modification by a factor of $\sqrt{f_{\pi\pi}(\pi_0)/2}$ is obtained in
the minisuperspace model if its quadratic dependence on the momentum $p$ is
replaced using the same function $f(p)$.

In holonomy-modified models of gravity, in which the quadratic
momentum-dependence is replaced by a bounded function $f$ just as in the model
used here, signature change is an interesting consequence
\cite{Action,SigChange,SigImpl}. This phenomenon was first noticed by an
analysis of space-time structures in the presence of a modified kinetic term,
as well as in equations of motion derived from such Hamiltonians
\cite{ScalarHol}. In these models, signature change to 4-dimensional Euclidean
space happens whenever the second derivative of the modification function is
negative, or around local maxima of the function. Instead of a discrete
classical signature parameter $\epsilon=\pm1$, the modified space(-time)
structures have signature given by $\beta(p)=\frac{1}{2}f_{pp}(p)$, which
appears in the new coefficient in (\ref{WIRHol}). When $\beta<0$, the infrared
contribution to the effective potential is imaginary, indicating an
instability. An instability is, in fact, the main consequence of modified
space-time structures with signature change, as analyzed in the cosmological
\cite{DeformedCosmo,SigImpl} and black-hole context \cite{Loss}.

While the full space-time structure cannot be seen in homogeneous
minisuperspace models, it is interesting to note that a minisuperspace model
of holonomy corrections is able to reproduce the correct factor in
(\ref{WIRHol}). However, the imaginary contribution to the potential means
that we have reached the limits of the adiabatic approximation, making it
difficult to interpret the complex value within the minisuperspace
setting. Nevertheless, the same conclusion of instabilities is then confirmed
by an analysis of equations of motion in the corresponding field theory, which
are elliptic rather than hyperbolic when $f_{pp}(p)/2<0$. We note that the
minisuperspace Hamiltonian is positive definite for any positive function
$f(p)$, even around a local maximum where $f_{pp}(p)<0$. We will discuss
signature change in more detail after we have developed further techniques in
the next section.

\section{Between minisuperspace and the full theory}
\label{s:Between}

The controlled setting of scalar (field) theories has allowed us to introduce
a minisuperspace approximation instead of just a truncation of degrees of
freedom: The quantum-mechanical effective potential can, up to a numerical
factor, be obtained from the field-theoretical Coleman--Weinberg potential by
restricting the latter to its infrared contribution and expanding the
resulting integral by $(\sqrt{W_{\phi\phi}(\phi_0)}L_0)^{-1}$, corresponding
to a potential term dominating the spatial-derivative term in the
Hamiltonian. One could use higher-order terms of such an expansion, as for
instance in (\ref{WIRCompact}), in order to go beyond the minisuperspace
result. However, one would have to begin with the full field-theory
calculation, so that this kind of approximation does not seem to lead to
strong simplifications. In this section, we explore the possibility of
introducing models which are simpler than the full theory but still capture
quantum terms that cannot be seen in a strict minisuperspace setting.

\subsection{Perturbative inhomogeneity}

Instead of truncating the canonical fields $\phi(x)$ and $\pi(x)$ to their
average values, as the starting point of a minisuperspace quantization, we
decompose the fields into their average values $\bar{\phi}$ and $\bar{\pi}$,
as well as their zero-average variations $\delta\phi$ and $\delta\pi$. (The
average values $\bar{\phi}$ and $\bar{\pi}$ will appear in some expressions in
a form very similar to $\phi_0$ and $\pi_0$ before. We use a different
notation for them in order to highlight their different origin in a
perturbative expansion, rather than just constant field values.) Such
decompositions have been used in canonical cosmological models going back to
\cite{Halliwell}, and more recently in \cite{ConstraintAlgebra,PertObsII}; see
also \cite{DeformedCosmo}. Formally, this decomposition can be introduced by
defining
\begin{equation}
 \bar{\phi}:= \frac{1}{L_0} \int_0^{L_0}{\rm d}x \phi(x) \quad,\quad \bar{\pi}:=
 \int_0^{L_0}{\rm d}x \pi(x)
\end{equation}
with integrations over the averaging volume of length $L_0$, and the
variations
\begin{equation}
 \delta\phi(x):=\phi(x)-\bar{\phi} \quad,\quad
 \delta\pi(x):=\pi(x)-\frac{1}{L_0}\bar{\pi}\,.
\end{equation}
(Anticipating the Poisson brackets derived below, we absorb a factor of $L_0$
in $\bar{\pi}$ in order to have canonical fields. Absorbing $L_0$ in
$\bar{\pi}$, rather than $\bar{\phi}$, mimics the density weight of the
corresponding field.)  By definition, the variations satisfy the conditions
\begin{equation} \label{Aver}
 \int_0^{L_0}{\rm d}x \delta\phi(x)=0=\int_0^{L_0}{\rm d}x \delta\pi(x)\,.
\end{equation}
They ensure that we do not double-count the average degrees of freedom,
which have been separated off as $\bar{\phi}$ and $\bar{\pi}$.

The canonical structure of decomposed fields follows from the original one,
for instance by pulling back the symplectic or Liouville form. For the average
values, computing
\begin{equation} \label{PB}
\int_0^{L_0}{\rm d}x \dot{\phi}(x)\pi(x)=
\dot{\bar{\phi}}\bar{\pi}+ \int_0^{L_0}{\rm d}x \delta\dot{\phi}(x)\delta\pi(x)
\end{equation}
using (\ref{Aver}), one sees that $\bar{\pi}$ is indeed the momentum of
$\bar{\phi}$, so that
\begin{equation}
 \{\bar{\phi},\bar{\pi}\} = 1
\end{equation}
for the average values. The variations are canonically conjugate, but also
subject to the second-class constraints (\ref{Aver}). Therefore, we switch
from the direct result of (\ref{PB}) to
their Dirac brackets
\begin{equation}
 \{\delta\phi(x),\delta\pi(y)\} = \delta(x,y)- L_0^{-1}\,.
\end{equation}
(With two second-class constraints $C_1$ and $C_2$ given in (\ref{Aver}), the
Dirac bracket is obtained by subtracting from the original Poisson bracket
the term $1/\{C_1,C_2\}=1/\int {\rm d}x \int{\rm d}y \delta(x,y)=1/L_0$.)

The Poisson brackets can be used to compute Hamiltonian equations of motion
from the decomposed Hamiltonian
\begin{eqnarray}
 H&=&\int_0^{L_0} {\rm d}x \left(\frac{1}{2}
   \pi(x)^2+\frac{1}{2}\phi'(x)^2+W(\phi)\right)\nonumber\\
 &=& \frac{\bar{\pi}^2}{2L_0} +L_0W(\bar{\phi}) + \frac{1}{2} \int_0^{L_0}
 {\rm d}x  \left(
   \delta\pi(x)^2+\delta\phi'(x)^2+ W_{\phi\phi}(\bar{\phi})
   \delta\phi(x)^2+\cdots\right)
\end{eqnarray}
where the dots now indicate terms of higher than second order in the
variations. By the decomposition of canonical fields and the expansion of $H$
to the given order, the original interacting field theory is converted into a
non-harmonic mechanical system coupled to a free field theory. Extending the
minisuperspace quantization, we can now quantize this system and obtain
non-harmonic quantum mechanics coupled to a free quantum-field theory. Such a
system is simpler than the full interacting quantum-field theory. Even if one
expands the Hamiltonian to higher orders in the variations, there may be
simplifications because a possibly non-polynomial field-theory potential would
be converted into a polynomial expansion.

Effective potentials of the decomposed theory can be computed using moments of
the two subsystems, $G^{c,d}$ for the quantum-mechanics part and
$G^{a,b}(x_1,\ldots,x_b;y_1,\ldots,y_a)$ for the field-theory part. As a
consequence of (\ref{Aver}), the field-theory moments are subject to the
conditions
\begin{equation}
 \int_0^{L_0} {\rm d}x G^{a,b}(\ldots,x,\ldots)=0\,.
\end{equation}
Their Poisson brackets follow using the Dirac bracket for variations, for
instance
\begin{eqnarray}
 \{G^{0,2}(x_1,x_2),G^{2,0}(y_1,y_2)\}&=& G^{1,1}(x_2,y_2)
 \left(\delta(x_1,y_1)-L_0^{-1}\right)+
 G^{1,1}(x_1,y_1)\left(\delta(x_2,y_2)-L_0^{-1}\right)\\
&& +G^{1,1}(x_2,y_1)\left(\delta(x_1,y_2)-L_0^{-1}\right)
 +G^{1,1}(x_1,y_2)\left(\delta(x_2,y_1)-L_0^{-1}\right)\,.\nonumber
\end{eqnarray}

Equations of motion for second-order moments take the same form as those
obtained for the full field theory. Therefore, the effective potential is the
same, except that the restricted space of the averaging volume requires us to
limit the wave numbers to be larger than $k_{\rm min}=2\pi/L_0$. Wave lengths
larger than the averaging volume are therefore excluded from the remaining
Coleman--Weinberg potential for the quantum field theory of variations, but
they are included in the effective potential of the quantum-mechanics part. Up
to numerical factors, the full Coleman--Weinberg potential is therefore split
correctly into the infrared contribution as in (\ref{Veff}), plus the
remainder.

As before, the numerical factors can be matched exactly in a periodic model
with its clearer separation of discrete modes. We have already seen that the
minisuperspace model produces exactly the infrared contribution to the
Coleman--Weinberg potential if we choose the averaging volume to equal the
length of the periodic space. The same term is now produced by the
quantum-mechanics part of our extended model. The field-theory part is almost
identical to the previous model, except that the conditions (\ref{Aver}) for
variations remove the zero-mode from the Fourier sum in
(\ref{Weffcompact}). But the zero mode, split off explicitly in
(\ref{WIRcompact}), is just the quantum-mechanics contribution, so that the
infrared contribution of our decomposed model is identical with the infrared
contribution in the full quantum-field theory, to the orders considered.

\subsection{Application: Signature change}

As a simple application of the perturbed model in the context of effective
potentials, we now revisit the question of signature change in
holonomy-modified theories. The quantum-mechanical minisuperspace model gives
a hint of signature change because the $\hbar$-correction to its effective
potential, with a factor of $\sqrt{f_{pp}(p)/2}$ if the quadratic momentum
dependence of the classical Hamiltonian is replaced by some function $f(p)$,
is imaginary around a local maximum of $f$, where $f_{pp}<0$. However, the
adiabatic approximation used to derive the effective potential breaks down in
this regime, so that the conclusion of unstable behavior is not reliable in
this setting. The minisuperspace model has a non-standard but positive
Hamiltonian and is well-defined. Moreover, the instability by itself does not
directly indicate signature change, even though field theories on Euclidean
space lead to instabilities of initial-value problems.

The perturbed model allows us to address the question without having to go to
the full field theory. We write the holonomy-modified Hamiltonian as
\begin{eqnarray}
 H&=&\int_0^{L_0} {\rm d}x
 \left(f(\pi(x))+\frac{1}{2}\phi'(x)^2+W(\phi)\right)\nonumber\\
 &=& \frac{\bar{\pi}^2}{2L_0} +L_0W(\bar{\phi}) +\frac{1}{2} \int_0^{L_0} {\rm d}x
 \left( f_{\bar{\pi}\bar{\pi}}(\bar{\pi})
   \delta\pi(x)^2+\delta\phi'(x)^2+W_{\phi\phi}(\bar{\phi})
   \delta\phi(x)^2+\cdots\right)\,.
\end{eqnarray}
It is now clear that the field-theory Hamiltonian of perturbative
inhomogeneity is not positive definite when
$f_{\bar{\pi}\bar{\pi}}(\bar{\pi})<0$. Already at the classical level, the
Hamiltonian equations
\begin{equation}
 \delta\dot{\phi} = f_{\bar{\pi}\bar{\pi}}(\bar{\pi}) \delta\pi \quad,\quad
 \delta\dot{\pi} =
 \delta\phi''- W_{\phi\phi}(\bar{\phi}) \delta\phi
\end{equation}
imply as a second-order field equation the mixed-type partial differential
equation
\begin{equation}
  -\delta\ddot{\phi} +f_{\bar{\pi}\bar{\pi}}(\bar{\pi}) \delta\phi''=
  f_{\bar{\pi}\bar{\pi}}(\bar{\pi})
  W_{\phi\phi}(\bar{\phi}) \delta\phi-
  \frac{f_{\bar{\pi}\bar{\pi}\bar{\pi}}(\bar{\pi})}{f_{\bar{\pi}\bar{\pi}}(\bar{\pi})}
\dot{\bar{\pi}}  \delta\dot{\phi}\,.
\end{equation}
For $f_{\bar{\pi}\bar{\pi}}(\bar{\pi})<0$, the equation is elliptic and is
well-posed with a 2-dimensional boundary-value problem instead of an
initial-value problem.

It is not necessary to include moment terms or consider a quantum-field theory
in order to see signature change, in contrast to the minisuperspace model. If
we include moments and compute the Coleman--Weinberg potential of the modified
(perturbed or full) theory, we obtain (\ref{WeffHol}). For
$f_{\bar{\pi}\bar{\pi}}(\bar{\pi})<0$, the $\hbar$-correction to the classical
potential has an imaginary factor, as it has it for a Euclidean quantum-field
theory with the same potential. Regarding signature change, the classical
perturbed model and the quantized perturbed or full theory therefore
agree. This observation supports the conclusions about signature change in
models of loop quantum gravity \cite{Action,SigChange,SigImpl}, which have
been made by considering spherically symmetric or perturbed cosmological
models without including moment terms \cite{JR,ScalarHol}.

\paragraph{Avoiding an adiabatic approximation.}

In order to discuss signature change in more detail, we go back to our
derivation of effective potentials in holonomy-modified models. In particular,
we should revisit the adiabatic assumption in this case because it is tied to
an evolution picture or initial-value problem, which is available only in the
Lorentzian regime but not in the Euclidean one if there is signature
change. It turns out that our previous Lorentzian solutions can be recovered
without using an adiabatic approximation, but they still rely on the presence
of a well-posed initial-value problem.

The three equations of motion for the second-order moments are given by
(\ref{G02F}), (\ref{G11SC}) and (\ref{G20SC}). Instead of applying the
adiabatic approximation, we eliminate $G^{1,1}$ to get the two equations
\begin{eqnarray}
\ddot{G}^{0,2}(y,z,t)&=&\frac{1}{2}f_{\pi\pi}\left[f_{\pi\pi}G^{0,2}(y,z,t)
  -\left(2W_{\phi\phi}-\frac{{\rm d}^2}{{\rm d}y^2}-\frac{{\rm d}^2}{{\rm
        d}z^2}\right)\right]
G^{0,2}(y,z,t)\\
\ddot{G}^{2,0}(y,z,t)&=&
-\left(W_{\phi\phi}-\frac{{\rm d}^2}{{\rm
      d}z^2}\right)\left[\frac{f_{\pi\pi}}{2}G^{2,0}(y,z,t)
  -\left(W_{\phi\phi} -\frac{{\rm d}^2}{{\rm d}y^2}\right)G^{0,2}(y,z,y) \right]
+(y\leftrightarrow z) \label{G20elim}\,.
\end{eqnarray}
We further assume that $\pi_0$ and $\phi_0$ in the coefficients
$f_{\pi\pi}(\pi_0)$ and $W_{\phi\phi}(\phi_0)$ are spatially constant. Then
eliminating $G^{2,0}$ from (\ref{G20elim}), we obtain a fourth-order
differential equation for $G^{0,2}$
\begin{eqnarray}
\left[\frac{{\rm d}^4}{{\rm d}t^4}
  -f_{\pi\pi}\frac{{\rm d}^2}{{\rm d}t^2}\left(\frac{{\rm d}^2}{{\rm d}y^2}
    +\frac{{\rm d}^2}{{\rm d}z^2} -2W_{\phi\phi}\right)
  +\frac{1}{4}f_{\pi\pi}^2\left(\frac{{\rm d}^2}{{\rm d}y^2} -\frac{{\rm
        d}^2}{{\rm d}z^2} \right)^2
\right] G^{0,2}(y,z,t)=0\,.
\end{eqnarray}

For $f_{\pi\pi}<0$, the principal symbol of this equation is positive
semidefinite, and therefore the equation is elliptic, while the equation has
characteristics for $f_{\pi\pi}>0$. In the latter case, it is therefore
meaningful to look for solutions determined by initial values at some fixed
$t$, which by Fourier decomposition can be written as
\begin{eqnarray} \label{G02In}
G^{0,2}(y,z,t)=\int {\rm d}k_y\int {\rm d}k_z\left[g_1(k_y,k_z)
  e^{i(k_yy+k_zz-\omega_k t)} + g_1^*(k_y,k_z) e^{-i(k_yy+k_zz-\omega_k
    t)}\right]\,,
\end{eqnarray}
with $\omega_k=\sqrt{f_{\pi\pi}/2}\left(\omega_y-\omega_z\right)$, where
$\omega_i=\sqrt{W_{\phi\phi} +k_i^2}$. Here we are choosing the minus sign so
that we get the correct limit for the usual Minkowski case, on which we will
comment further below. (Solutions to the fourth-order differential equation
would be consistent with a sum $\omega_y+\omega_z$ as well.) Since we assume
$f_{\pi\pi}>0$ for now, $\omega_k$ is real. For $f_{\pi\pi}<0$, only imaginary
solutions for $\omega_k$ exist, consistent with the absence of characteristics
in this elliptic case.

The arbitrariness of
the state about which these functions are evaluated is captured by the
function $g_1(k_y, k_z)$ and its conjugate. These can be fixed by specifying
the initial conditions for the moments through the relations
\begin{eqnarray}
g_1(k_y,k_z)&=&\frac{1}{2}\left[G^{0,2}(k_y,k_z)+
\frac{i}{\omega_k}\dot{G}^{0,2}(k_y,k_z)\right]\\
g_1^*(k_y,k_z)&=&\frac{1}{2}\left[G^{0,2}(-k_y,-k_z)-
\frac{i}{\omega_k}\dot{G}^{0,2}(-k_y,-k_z)\right]\,,
\end{eqnarray}
where the newly introduced function $G^{0,2}(k_y,k_z)$ and its time derivative
are defined as
\begin{eqnarray}
G^{0,2}(k_y,k_z):=\int {\rm d}y\int {\rm d}z \,\,
G^{0,2}(y,z,t=0)e^{-i[k_yy+ k_zz]}\\
\dot{G}^{0,2}(k_y,k_z):=\int {\rm d}y\int {\rm d}z \,\,
\dot{G}^{0,2}(y,z,t=0)e^{-i[k_yy +k_zz]}\,.
\end{eqnarray}

Further, if the theory is assumed to have translational invariance in space,
we can use $g_1(k_y, k_z)=\tilde{g}(k_y)\delta(k_y+k_z)$ to get
\begin{eqnarray}
G^{0,2}(y,z,t)=\int {\rm d}k \left[\tilde{g}(k) e^{ik(y-z)} + \tilde{g}^*(k)
  e^{-ik(y-z)}\right]\,.
\end{eqnarray}
We can immediately see that these moments do not depend on time explicitly,
even without using an adiabatic approximation. Having a (spatially)
translationally invariant theory is sufficient to imply time translation
invariance in the Lorentzian case. Physically, translation invariance in space
implies the absence of propagating modes in initial values, so that time
translation invariance of solutions is implied on a static background.  While
this heuristic interpretation justifies the conclusion about symmetric
solutions, it follows only because $\omega_k$ has the form given after
Eq.~(\ref{G02In}), which is a direct consequence of the form of the equations
of motion, {\em together with} the sign choice commented on above. One could
therefore invert these arguments and fix the sign choice
$\omega_k=\sqrt{f_{\pi\pi}/2}\: (\omega_y-\omega_z)$, as opposed to
$\omega_k=\sqrt{f_{\pi\pi}/2}\: (\omega_y+\omega_z)$, avoiding a comparison
with the usual Minkowski solutions.

If we further demand that the theory have reflection symmetry, that is
$\tilde{g}(-k) =\tilde{g}(k)$, it is evident that we can write the
solution of $G^{0,2}(y,z)$ as
\begin{eqnarray}
 G^{0,2}(y,z) = \int {\rm d}k g(k) e^{ik(y-z)}\,,
\end{eqnarray}
where we have absorbed a factor of $2$ in the function $g(k)$ and have hence
dropped the tilde. Thus we have the same solution for the moments, even
without the adiabatic approximation, as long as we have spatial translational
invariance and reflection symmetry. Both of these properties are expected to
be realized even with modified momentum dependence (motivated by some kind of
spatial quantum geometry) in a continuum effective theory.

For $f_{\pi\pi}<0$, the initial-value formulation based on (\ref{G02In}) with
$g_1(k_y,k_z)$ related to the values of $G^{0,2}$ at some fixed time, should
be replaced by a boundary-value problem in the $t$-direction. We then have to
prescribe $G^{0,2}(y,z,t)$ at two fixed values of $t=t_1$ and $t=t_2$. It is
then clear that time translation invariance can be respected by solutions only
in the special case in which we choose
$G^{0,2}(y,z,t_1)=G^{0,2}(y,z,t_2)$. There seems to be no independent physical
condition that justifies such a choice, which is consistent with the formal
result that the adiabatic approximation breaks down for $f_{\pi\pi}<0$: In
this case, using the adiabatic approximation results in imaginary
contributions to the effective potential. While a complex potential, or
complex moments obtained in the derivation of such a potential, are not
meaningful, imaginary contributions can be interpreted as indications of
instabilities which occur when one attempts to solve elliptic partial
differential equations by initial-value problems.

\paragraph{Euclidean theories.}

In our discussion of holonomy modifications, we have seen that a change of
sign in $f_{\pi\pi}$ implies that second-order moments and the effective
potential acquire an imaginary factor. However, $G^{0,2}$ and $G^{2,0}$ are
fluctuations and should never be negative, let alone be imaginary. We can
resolve this apparent inconsistency by comparing the effective
holonomy-modified theory with standard Euclidean field theory, motivated by
the observations made elsewhere that a change of sign in $f_{\pi\pi}$
indicates signature change. Such a comparison will allow us to elucidate the
role of signature change further.

We begin by recalling that the Coleman--Weinberg potential takes the same form
in Lorentzian and Euclidean quantum-field theory, as stated already in
\cite{ColemanWeinberg}. Some details leading to this result will be relevant
for our further comments on signature change: The Euclidean action (with one
spatial dimension as before) is traditionally defined as
\begin{equation}
 S_{\rm E} = \int{\rm d}t {\rm d}x \left(\frac{1}{2} \left(\frac{{\rm
         d}\phi}{{\rm d}t}\right)^2+ \frac{1}{2}\left(\frac{{\rm d}\phi}{{\rm
         d}x}\right)^2+ W(\phi)\right)
\end{equation}
with a positive potential term. With this choice, the Euclidean Hamiltonian
\begin{equation}
 H_{\rm E} = \int{\rm d}x \left(\frac{1}{2}\pi^2- \frac{1}{2}
   \left(\frac{{\rm d}\phi}{{\rm d}x}\right)^2-W(\phi)\right)
\end{equation}
is transformed to the positive-definite energy functional
\begin{equation}
 -H_{\rm EW} = \int{\rm d}x \left( \frac{1}{2}\left(\frac{{\rm d}\phi}{{\rm
         d}\tau}\right)^2+ \frac{1}{2} \left(\frac{{\rm d}\phi}{{\rm
         d}x}\right)^2+W(\phi)\right)
\end{equation}
after a Wick rotation from $t$ to $\tau=it$ (so that $\pi={\rm d}\phi/{\rm d}
t= i{\rm d}\phi/{\rm d}\tau$. The weight $\exp(iS)$ of a Lorentzian path
integral then becomes the correct weight $\exp(\int{\rm d}\tau H_{\rm
  EW})=\exp(-\int{\rm d}\beta E)$ of the partition function of statistical
mechanics, with the energy $E$ (and a periodic range for $\beta$). Since
$H_{\rm EW}$ is, up to a total minus sign and the use of $\tau$ instead of $t$,
the same as the Hamiltonian of the Lorentzian theory, our canonical methods
imply the same Coleman--Weinberg potential independently of the signature, in
agreement with \cite{ColemanWeinberg}.

Signature change in models of loop quantum gravity leads to different results
because they are not accompanied by a Wick rotation. The momentum term in the
Hamiltonian changes sign, which leads to elliptic field equations as on
Euclidean space, but all coordinates remain real.  The Hamiltonian obtained
with holonomy modifications has a non-quadratic kinetic term $f(\pi(x))$,
which remains positive so that in this form we do not see directly what role
the sign of $f_{\pi\pi}$ should play. We can make this coefficient appear
explicitly if we consider a Hamiltonian expanded for small variations
$\delta\phi$ and $\delta\pi$ around some background fields $\phi_0$ and
$\pi_0$, which would be constant when used in the Coleman--Weinberg potential:
\begin{eqnarray}
 H&=&\int{\rm d}x
 \left(f(\pi_0)+\frac{1}{2}(\phi_0')^2+W(\phi_0)\right)\nonumber\\
 && +\int{\rm d}x
 \left(\frac{1}{2} f_{\pi\pi}(\pi_0)
   \delta\pi(x)^2+\frac{1}{2}\delta\phi'(x)^2+\frac{1}{2}W_{\phi\phi}(\phi_0)
   \delta\phi(x)^2+\cdots\right)\,.
\end{eqnarray}
Up to a total minus sign, the Hamiltonian for perturbations, with
$f_{\pi\pi}(\pi_0)=-1<0$, is identical to the Euclidean Hamiltonian $H_{\rm
  E}$, but without any Wick rotation the Hamiltonian remains of indefinite
sign and is unbounded from above and below.

Perturbative modes evolve according to a Hamiltonian in which the sign of the
kinetic term is given by the sign of $f_{\pi\pi}$. Signature change is implied
because the kinetic and spatial-derivative terms have different relative signs
depending on the sign of $f_{\pi\pi}$, and correspondingly, second-order field
equations for $\delta\phi$ are hyperbolic partial differential equations for a
positive $f_{\pi\pi}$, but elliptic ones for a negative $f_{\pi\pi}$. The
relative signs of these two terms therefore agree with what one obtains for
field theories on Lorentzian (positive $f_{\pi\pi}$) and Euclidean space-times
(negative $f_{\pi\pi}$), respectively, provided these space-times are kept
real and not modified by Wick rotations. The non-positive Hamiltonian then
leads to instabilities, as shown by imaginary quantum corrections to effective
potentials.

\section{Conclusions}

In a variety of scalar field theories, we have found good qualitative
agreement between infrared contributions to their Coleman--Weinberg-type
potentials and effective potentials in quantum-mechanical systems. With simple
periodic spaces, quantitative agreement was found. These models provide
several instructive conclusions about the question how minisuperspace
truncations can be related to field theories, and how to go beyond the
minisuperspace setting in a controlled way.

\subsection{Minisuperspace truncation vs.\ approximation}

In quantum cosmology, minisuperspace models have been studied for decades,
after their introduction in \cite{DeWitt}; see also \cite{OUP}. While their
relation to some putative full theory remains weak, it is has been known for
some time that certain dynamical properties of the full theory cannot be
captured in this setting \cite{MiniValid}. More recently, signature change has
joined the list of concrete phenomena which, at high density or large momenta,
can have important implications not seen in minisuperspace truncations.

Loop quantum cosmology \cite{LivRev,Springer} was initially motivated by the
hope that the controlled kinematical setting of loop quantum gravity might
make it possible to derive some aspects of reduced models
\cite{SymmRed}. However, also in this context, not much progress on strict
derivations of the dynamics has been made, even setting aside the problem that
the dynamics of full loop quantum gravity remains poorly controlled owing to
quantization ambiguities and possible anomalies. (Some progress on the latter
question has recently been made
\cite{TwoPlusOneDef,TwoPlusOneDef2,AnoFreeWeak}, but so far without
indications on the minisuperspace question.) The kinematical side of a
possible reduction of loop quantum cosmology from loop quantum gravity is
still being analyzed
\cite{AinvinA,AsympAlmostPeriodic,SymmStatesInt,NonLinEmb,InvConn,ProjLQC,InvChar,KinLQC},
and possible dynamical relations between minisuperspace models and the full
theory are tentative. A recent idea uses condensate states in order to
describe homogeneous space-times, either as an approximation \cite{NonLinLQC}
or as a reduction from a full theory defined by group-field theories
\cite{GFTCosmo,GFTCosmo2,GFTCosmo3,GFTEx,GFTLQC,GFTLattice,GFTPerturb}. Symmetry
reduction of standard quantum-field theories has been used in \cite{SymmQFT}
in order to understand minisuperspace truncations in a controlled setting, but
so far only for free theories to which the conclusions of \cite{SymmQFT}
appear to be restricted. In the present paper, we have provided results for
interacting quantum-field theories, focussing on effective potentials in order
to overcome the more-complicated questions of how to relate states.

In our models, a quantum-mechanical system does not just appear as a
minisuperspace truncation of a field theory with a reduced number of degrees
of freedom, but can be embedded in a controlled approximation. The infared
contribution to a Coleman--Weinberg potential is seen to agree, up to a
numerical factor, with the quantum-mechanical effective potential to leading
order in an expansion by $(\sqrt{W_{\phi\phi}(\phi)}L_0)^{-1}=(2\pi)^{-1}
k_{\rm IR}/\sqrt{W_{\phi\phi}(\phi)}$ with the infrared scale $L_0$ or wave
number $k_{\rm IR}$. (For a periodic space with discrete wave numbers, we have
found exact agreement with the quantum-mechanics result.)

Taken to higher orders, this expansion corresponds to an approximation with
controlled correction terms, valid as long as the potential dominates the
spatial-derivative term in the Hamiltonian. There is an important difference
between traditional minisuperspace truncations and the minisuperspace
approximation provided here: While truncations are usually performed at the
kinematical level, introducing the dynamics by formulating a quantum
Hamiltonian on the state space of the reduced model, our minisuperspace
approximation is dynamical in a crucial way. We need to solve some of the
field-theory equations of motion for moments in order to obtain the effective
potential whose infrared contribution we expand. The leading order of this
expansion can be derived in a truncated model, just using the classical
potential and moment equations of quantum mechanics. Going to the next orders
is then not just a matter of computing a higher-order term, but requires
more-detailed information about the full dynamics. In this sense, there is a
big leap between minisuperspace truncations and the approximations developed
here.

It might seem pointless to compute a minisuperspace approximation if it
requires one to solve the full dynamical equations first. At this stage, it is
important that, at least to first order in $\hbar$, we can derive the expanded
infrared contribution to the effective potential by combining the
semiclassical approximation with an expansion in terms of perturbative
inhomogeneity as shown in Sec.~\ref{s:Between}. The resulting models between
minisuperspaces and the full theory take the form of quantum mechanics coupled
to a free quantum-field theory (again, to first order in $\hbar$), in which
semiclassical equations of motion for moments are manageable. The formal
setting is closely related to canonical cosmological perturbation theory.

\subsection{Infrared scale}

The minisuperspace correspondence found here explains the meaning of the
dependence of quantum corrections on the averaging volume $V_0$, or the
averaging distance $L_0$ in our 1-dimensional models. The situation within
minisuperspace models has been very unclear in the context of quantum
cosmology, in which the physical meaning of $L_0$ is hard to see. Moreover,
while physical results in the classical theory do not depend on the value of
$L_0$, quantum corrections usually do. We have explained this discrepancy by
the fact that classical theories are local, while quantum theories have
non-local features so that they can be sensitive to averaging volumes via the
number of modes included in quantum corrections. Moreover, we have given the
parameter $L_0$ a direct physical meaning as the infrared scale of
quantum-field theory. Contributions to effective potentials depend on the
number of modes included in their derivation, so that a dependence of quantum
corrections on $L_0$ is nothing but the well-known running of coupling
constants in quantum-field theory.

Our results also show that $L_0$ is not an infrared cut-off or a regulator, as
sometimes suggested in loop quantum cosmology; see for instance
\cite{InfReg}. The limit $L_0\to\infty$ is not required in order to have
qualitative agreement of minisuperspace and quantum-field theory results; this
limit would just eliminate all modes considered for the effective potential,
and therefore lead to a vanishing contribution without any qualitative
comparison. In fact, in the models considered, no infrared cut-off is required
because the effective potentials are infrared finite. If one computes the
Coleman--Weinberg potential by summing up 1-loop Feynman diagrams with a free
number of external lines, as introduced in \cite{ColemanWeinberg}, one may
encounter infrared divergences: For the original $\lambda\phi^4$-potential,
\begin{equation}
 W_{\rm eff}(\phi) = \lambda\phi^4+ \frac{\hbar}{(2\pi)^4}\int {\rm d}^4k
 \sum_{n=1}^{\infty} \frac{1}{2n}
 \left(\frac{12\lambda\phi^2}{k^2+i\epsilon}\right)^n
\end{equation}
has infinite contributions if each term in the sum (for $n>1$) is integrated
individually at small $k$.  However, these infrared divergences disappear in
the final effective potential if the sum is performed before integrating,
giving rise to (\ref{CW}). While the ultraviolet divergence remains, the
infrared one is replaced by a logarithmic divergence of the effective
potential at $\phi_0=0$. The infrared contributions used here do not encounter
this last divergence because it is outside of the range of the expansion by
$(\sqrt{W_{\phi\phi}(\phi_0)}L_0)^{-1}$. Therefore, a minisuperspace
approximation of scalar quantum-field theory does not require an infrared
cut-off.

Unlike an ultraviolet cut-off, the infrared scale used here does not depend on
unknown physics. It has a clear physical meaning as a selection of modes
included in the averaged effective potential. Therefore, it is sufficient to
work with a given scale $L_0$, once selected, and no renormalization is
required. One could try to absorb $L_0$ completely in renormalized coupling
constants, as done for renormalizable theories with the ultraviolet
cut-off. However, it is easy to see that this can be possible only for a few
special potentials. To first order in $\hbar$, $\sqrt{W_{\phi\phi}(\phi_0)}$
would have to be of the same functional form as $W(\phi_0)$ if a combined
coupling constant could be used in $W(\phi_0)+\frac{1}{2} \hbar
\sqrt{W_{\phi\phi}(\phi_0)}/L_0$ that absorbs $L_0$. The potential would have
to satisfy the differential equation $W_{\phi\phi}(\phi)=AW(\phi)^2$ with an
arbitrary constant $A$. This non-linear second-order equation can be
transformed to two coupled first-order ones by introducing $U:=W_{\phi}$, so
that $U {\rm d}U/{\rm d}W=W_{\phi}{\rm d}U/{\rm
  d}W=U_{\phi}=W_{\phi\phi}=AW^2$ or, solving $U{\rm d}U=AW^2{\rm d}W$,
$W_{\phi}=\sqrt{\frac{2}{3}AW^3+c}$ with another constant $c$. The remaining
equation can be integrated by separation, but the resulting expression for $W$
has a closed inverse only in the case $c=0$, for which $W(\phi)=a\phi^{-2}$
with a constant $a$. In this case, the effective potential can be written as
\begin{equation}
 W_{\rm eff}(\phi_0)=\frac{a+\sqrt{6a} \hbar/L_0}{\phi_0^2}=: \frac{a_{\rm
     ren}}{\phi_0^2}
\end{equation}
with a renormalized $a_{\rm ren}$. If the bare coupling constant $a$ is
allowed to depend on $L_0$, $a_{\rm ren}$ may be assumed to be independent of
$L_0$ and a scale-free theory is obtained \cite{JackiwConformal}. However, potentials not obeying the
equation $W_{\phi\phi}\propto W^2$ do not give rise to scale-free effective
potentials.

\subsection{Quantum-geometry modifications}

Although we did not consider quantum gravity, our models can be used to show
that modifications suggested by quantum geometry, mainly loop quantum gravity
in the canonical setting used here, can be captured by corresponding
modifications in the minisuperspace models. The field-theory modifications
modelling discrete quantum space usually depend on microscopic parameters, for
instance the discreteness scale $\ell_0$. The minisuperspace result matches
the infrared contribution to the field-theory effective potential if the same
scale is used in minisuperspace modifications (rather than the only scale,
$L_0$, present in the minisuperspace model). In loop quantum cosmology, such a
dependence has been recognized in the context of lattice refinement
\cite{InhomLattice,CosConst}.

We have even seen hints of signature change in a minisuperspace model by way
of an instability implied by a complex effective potential. However, the
phenomenon itself lies outside of the range of validity of the adiabatic
approximations used. In order to make to effect reliable, one has to go beyond
the minisuperspace setting, as done in Sec.~\ref{s:Between} by introducing
perturbative inhomogeneity as it has been used in cosmological models of loop
quantum gravity as well. We have further elucidated signature change in models
of loop quantum gravity by contrasting the Euclidean versions implied by this
phenomenon with the usual notion of Euclidean quantum-field theory. A crucial
difference is the absence of a Wick rotation in the former case, so that
instabilities cannot be fully removed.

The scalar field theories studied here can model several aspects of features
expected for quantum gravity and cosmology. Since our new minisuperspace
approximation makes use of infrared contributions to effective potentials, the
non-renormalizability of perturbative quantum gravity is not an
issue. However, new ingredients would have to be included for a direct
application to quantum cosmology: First, gravitational models are necessarily
constrained systems, which requires an extension of the canonical effective
framework used here. Methods for effective constraints have been developed in
\cite{EffCons,EffConsRel}. Secondly, in the absence of an absolute time
parameter, it is not clear how the adiabatic expansion can be formulated. This
question can be circumvented by considering moment couplings to expectation
values without trying to solve for the moments separately. The emphasis would
then not be on effective potentials but on other suitable properties referring
more directly to the dynamics of moments.  Finally, one must address the
question of how an infrared scale can meaningfully be fixed in a
diffeomorphism-covariant theory.

\section*{Acknowledgements}

This work was supported in part by NSF grant PHY-1307408.


\newcommand{\noopsort}[1]{}

\end{document}